# Geometric-phase microscopy for quantitative phase imaging of isotropic, birefringent and space-variant polarization samples

Petr Bouchal,[1,2,*] Lenka Štrbková,[2] Zbyněk Dostál,[1,2] Radim Chmelík[1,2] and Zdeněk Bouchal[3]

[1]*Institute of Physical Engineering, Faculty of Mechanical Engineering, Brno University of Technology, Technická 2, 616 69 Brno, Czech Republic*
[2]*Central European Institute of Technology, Brno University of Technology, Purkyňova 656/123, 612 00 Brno, Czech Republic*
[3]*Department of Optics, Palacký University, 17. listopadu 1192/12, 771 46 Olomouc, Czech Republic*
*\*petr.bouchal@vutbr.ceitec.cz*

**Abstract:** We present geometric-phase microscopy allowing a multipurpose quantitative phase imaging in which the ground-truth phase is restored by quantifying the phase retardance. The method uses broadband spatially incoherent light that is polarization sensitively controlled through the geometric (Pancharatnam-Berry) phase. The assessed retardance possibly originates either in dynamic or geometric phase and measurements are customized for quantitative mapping of isotropic and birefringent samples or multi-functional geometric-phase elements. The phase restoration is based on the self-interference of polarization distinguished waves carrying sample information and providing pure reference phase, while passing through an inherently stable common-path setup. The experimental configuration allows an instantaneous (single-shot) phase restoration with guaranteed subnanometer precision and excellent ground-truth accuracy (well below 5 nm). The optical performance is demonstrated in advanced yet routinely feasible noninvasive biophotonic imaging executed in the automated manner and predestined for supervised machine learning. The experiments demonstrate measurement of cell dry mass density, cell classification based on the morphological parameters and visualization of dynamic dry mass changes. The multipurpose use of the method was demonstrated by restoring variations in the dynamic phase originating from the electrically induced birefringence of liquid crystals and by mapping the geometric phase of a space-variant polarization directed lens.

## Introduction

The dynamic phase of light directly related to the optical path has always played a key role in optics. The controlled change of the dynamic phase has become the basis for light shaping methods implemented by traditional optical components. The spatial changes of the dynamic phase imposed on light have been successfully utilized in quantitative phase imaging (QPI) and phase-sensitive measurement to observe weakly absorbing biological samples and to measure surface topography.[1,2,3] In current optics, controlling light through the geometric phase also becomes important thanks to emerging technology known as fourth-generation (4G) optics.[4,5,6] Exceptional features of 4G optics open entirely new pathways for developing various branches of optics, represented here by the multipurpose polarization sensitive QPI.

The QPI is a rapidly developing area of optics that excels by an ability to map variations of the optical path difference (OPD) caused by the light interaction with phase objects being examined.[7,8] In optical microscopy, a variety of advanced QPI techniques has recently appeared, opening entirely new possibilities for quantitative label-free live cell studies in life sciences.[9] Depending on the complexity of the phase encoding in the intensity records, the QPI can be loosely divided into interference and computational methods. In interference methods, the phase information is obtained from interference patterns created by sample and well-defined reference waves using procedures known from digital holography and interferometry[10] that can be combined with tomography algorithms.[11,12] The typical computational approaches to QPI include propagation-based methods using the transport of intensity equation[13] or techniques of conventional and Fourier ptychographic microscopy utilizing the phase-retrieval algorithms.[14] The available QPI methods differ in optical performance and implementation complexity, and each specific method has the strengths and weaknesses that need to be considered for potential applications. Widely used common-path QPI techniques[15] have an exceptional temporal stability resulting in a high measurement precision but provide the phase associated with the image field[16] replacing the ground-truth OPD.[17] These methods are based on the temporal coding, hence the phase is not restored instantaneously, and tracking of dynamic motions and morphological changes of cells is impossible. The ground-truth OPD is available in double-path QPI systems that work in real time.[18] Unfortunately, high vibration sensitivity and a very complex design make it difficult to use these systems in common operating conditions.

In the conventional QPI methods, the dynamic phase influenced by spatial changes of the sample refractive index is quantitatively restored using light shaping optical components also modulating the dynamic phase of light. Here, we present 4G optics platform for the polarization sensitive QPI employing the geometric phase associated with the evolution of light in an anisotropic parameter space. Although the geometric phase was discovered in quite distant history by Pancharatnam[19] and later widely elaborated by Berry,[20] the pioneering 4G optics technology enabling realization of unique geometric-phase elements has emerged just recently.[4] The physically thin and highly efficient elements of 4G optics are capable of a wavelength independent modulation of the geometric phase carried out by the polarization transformation. The most important property of the geometric-phase elements is their polarization selectivity allowing to impose phase changes of the opposite sign on the light with left-/right-handed circular polarization (LHCP/RHCP).[5,6] We have used this strategy to design experiments with optical performance unavailable in state-of-the-art equipment and to develop a geometric-phase QPI modality capable of quantifying spatial variations of the phase retardance between orthogonally polarized waves. The geometric-phase QPI platform, here referred to as quantitative 4G optics microscopy (Q4GOM), was developed as an incoherent achromatic imaging technique allowing the instantaneous (single-shot) restoration of the phase retardance on either dynamic or geometric phase. This capability makes Q4GOM powerful and versatile. The phase restoration uses self-interference of light and is implemented in an inherently stable common-path setup. The experiments are based on the use of an add-on 4G optics imaging module that is simply connected to either standard or polarization modified optical microscope. The multipurpose use of Q4GOM is demonstrated by quantitative noninvasive imaging of live cells, restoration of the dynamic phase retardance originating from birefringence of liquid crystals and complete quantitative mapping of geometric-phase holograms.[5]

## Methods

Optical setup and strategy of experiments

Q4GOM was implemented in the experimental configuration shown in Fig. 1. The key part of the setup is an add-on 4G optics imaging module capable of quantifying the phase retardance between orthogonally polarized waves. The polarized waves entering the module carry information on the sample phase and provide the pure (sample independent) reference phase guaranteeing high ground-truth accuracy of the measurement. Depending on the specific application, the add-on 4G optics module is used in two different ways. In experiments with isotropic samples, the add-on 4G optics module is connected to optical microscope using a polarization adapted imaging path, in which the signal and reference waves with orthogonal polarizations are prepared. When restoring the retardance of polarization sensitive samples, the add-on 4G optics module is connected to a conventional optical microscope, working either in reflection or transmission mode.

In the live cell imaging, the add-on 4G optics module was connected to a polarization adapted interference microscope (Fig. 1a, Supplementary material, Part A). The orthogonally polarized sample and reference waves were created by the Mirau microscope objective MO (Nikon 10x, NA=0.3) using the beam splitter $BS_1$ (Fig. 1b, Supplementary material, Part B). To operate the Mirau MO in a polarization mode, $BS_1$ was supplemented by the quarter-wave plates $QWP_1$ and $QWP_2$ with the directions of the anisotropy axis turned by 45° (details about design and adjustment of the polarization adapter are available in Supplementary material, Part B). The illumination path was created using a tungsten-halogen lamp with bandpass filter (central wavelength 600 nm, full width at half maximum 50 nm) and lenses $IL_1$ and $IL_2$ providing Köhler illumination conditions. The light entering the MO through the input polarizer $P_1$ and the beam splitter cube $BS_2$ is linearly polarized in the direction coinciding with the azimuth of the compensating quarter-wave plate $QWP_1$.

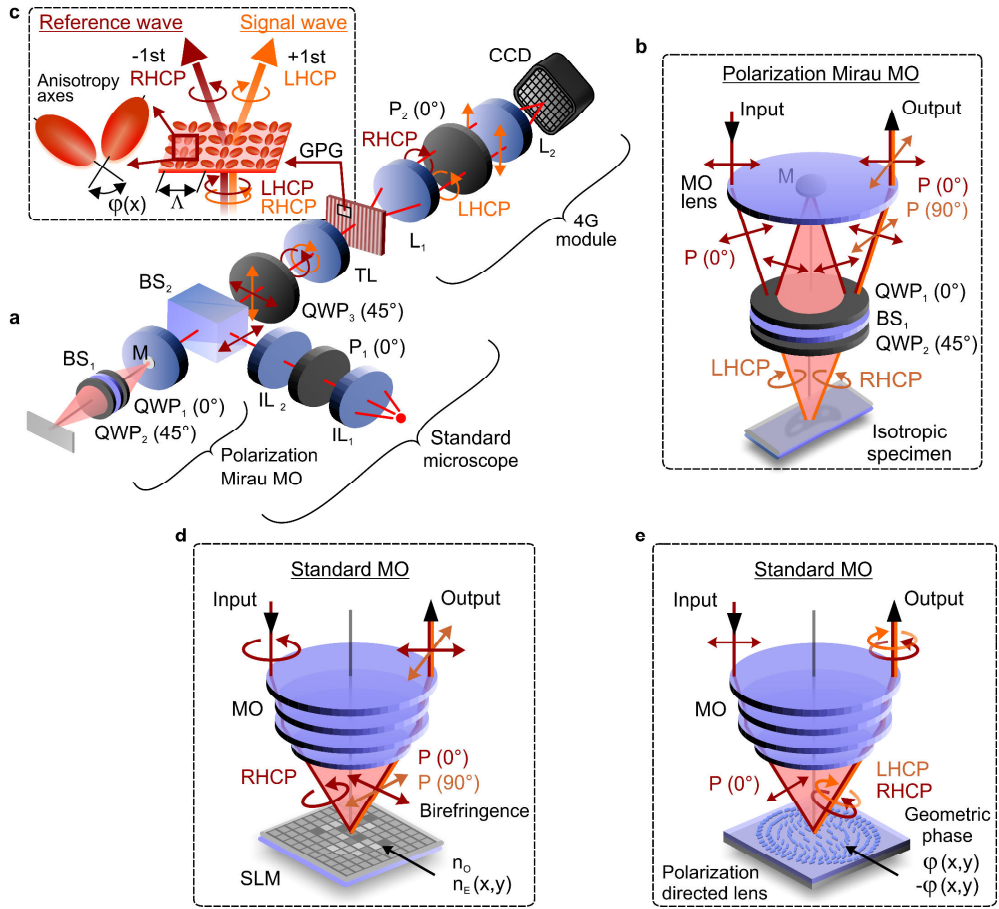

Fig. 1 Illustration of Q4GOM in experiments for quantification of the phase retardance of biological and polarization sensitive samples. (a) Experimental setup using 4G optics module connected to microscope with a polarization adapted interference objective: $P_1$–input polarizer, $IL_1$, $IL_2$–illumination lenses, MO–Mirau microscope objective, $BS_1$–pellicle beam splitter, $QWP_1$, $QWP_2$, $QWP_3$–quarter-wave plates, M–reference mirror, $BS_2$–beam splitter cube, TL–tube lens, GPG–geometric-phase grating, $L_1$–first Fourier lens, $P_2$–analyzer, $L_2$–second Fourier lens and CCD–charged coupled device. (b) Polarization adapted Mirau MO used for imaging of isotropic samples. (c) Polarization sensitive transformation of light by geometric-phase grating. (d) Polarization coded waves in restoration of the birefringent retardance of the dynamic phase introduced by liquid crystals. (e) Polarization coded waves in the quantitative mapping of the geometric phase modulated by multi-functional polarization directed elements.

Hence, the waves reflected from the sample and the reference mirror M of the MO get the orthogonal linear polarizations after double passing through $QWP_1$ and $QWP_2$ (Fig. 1b). Using the quarter-wave plate $QWP_3$, the orthogonal linear polarizations are transformed into LHCP and RHCP. The light collimated by the MO is focused by the tube lens TL (Nikon CFI60). The back focal plane of the TL coincides with the input plane of the add-on 4G optics module, where LHCP and RHCP images created in the sample and reference path overlap, and a polarization directed geometric-phase grating (GPG) is placed. The GPG was custom-made by Boulder Nonlinear Systems as a polymer liquid crystal grating with the spatially modulated orientation of the anisotropy axis. The spatial period corresponding to $2\pi$ rotation of the anisotropy axis was chosen as $\Lambda = 9\,\mu m$ (functionality of the GPG is discussed in detail in Supplementary material, Part C). When the sample and reference waves pass through the GPG, their polarization state is transformed from LHCP to RHCP and vice versa, while the geometric phase varies as $\pm 2\varphi$, where $\varphi$ defines the periodic spatial change of the angular orientation of the anisotropy axis (Fig. 1c). Because of the geometric-phase modulation, the sample and reference waves with the orthogonal circular polarizations are inclined to directions of +1 st and -1 st diffraction order with the mutual angle of $8°$ for the central wavelength. Although the modulation of the geometric phase is achromatic, the light propagation causes a diffractive dispersion whose manifestation is observable at the back focal plane of the Fourier lens $L_1$. Utilizing the separation of the spatial spectrum and the polarization coding of the sample and reference waves, the interferometric paths can advantageously be aligned by inserting a variable phase shifter to the back focal plane of the lens $L_1$. With this adjustment, the Mirau MO can be exploited for observing biological samples with a cover glass even when light with a short coherence length is used. After the polarization projection and the Fourier transformation carried out using the polarizer $P_2$ and the lens $L_2$, respectively, the interference field formed by the sample and reference waves is obtained on the CCD (Ximea MR4021MC-BH).

Besides demonstrated QPI of live cells, Q4GOM holds a great promise for quantitative measurement of the retardance on both dynamic and geometric phase created by birefringent samples, geometric-phase elements, plasmonic metasurfaces and chiral samples. Light interacting with these objects becomes naturally encoded in polarization, hence the add-on 4G optics module is connected to optical microscope using a standard MO. Here, these experiments are represented by restoration of the phase retardance originating from the birefringence of liquid crystals and from the rotation of anisotropy axis in polarization directed 4G optics elements.

Measurement of the birefringent liquid crystal retardance was carried out in the setup shown in Fig. 1, in which the liquid crystal display of a reflective spatial light modulator (SLM) was used as a testing sample and placed in the front focal plane of a standard MO (Fig. 1d). The SLM was illuminated by polarized light (diagonal linear polarization, circular polarization) formed by orthogonal linear polarizations. Using the birefringence of liquid crystals, the dynamic phase of one polarization component was modulated by controlling the refractive index of individual pixels, while the orthogonal polarization component was reflected without phase modulation. By a quarter-wave plate, the sample and reference waves with the orthogonal circular polarizations were created in the input plane of the 4G optics module.

The quantification of the geometric-phase retardance was demonstrated using a thin polarization directed lens positioned in the front focal plane of the MO and illuminated by linearly polarized light composed of orthogonal circular polarizations (Fig. 1e). The functionality of the lens is the same as that explained for the GPG (Supplementary material, Part C). The handedness of the circular polarizations is reversed by the lens, and in this transformation, the waves receive a change of the geometric phase, which differs in the sign (i.e. the waves are complex conjugate). Hence, the geometric-phase lens is polarization directed and acts as converging and diverging lens on the waves with the orthogonal circular polarizations. In Q4GOM, the spatially varying retardance of the geometric phase introduced by the polarization directed lens was measured using the add-on 4G optics module connected to a standard optical microscope.

Theoretical framework

Restoration of the phase retardance in Q4GOM is realized using broadband spatially incoherent illumination and involves transformations of the polarization and the geometric phase of light. The phase restoration is clarified in simulation model working with monochromatic components of the temporal Fourier spectrum of light illuminating the sample under Köhler conditions. The description of the polarization modification of the imaging path in measurement of isotropic samples is made using Jones formalism while the imaging of the sample into the input plane of the 4G module is examined by the convolution approach. The GPG performing the polarization transformation resulting in a spatially variable change of the geometric phase is described by Jones matrix acting in the basis of circularly polarized waves. The entire intensity record is described as a continuous superposition of the correlation patterns that originate from the interference of diffraction spots representing optically conjugated point images created in the sample and reference imaging paths. Because the individual correlation patterns are recorded with the achromatically created carrier frequency, the phase retardance is quantitatively reconstructed from a single shot by processing the space-time coherence function. Exact restoration of the phase retardance is demonstrated for fully incoherent illumination of the sample and an ideal point imaging. The detailed computational model of Q4GOM is elaborated in Supplementary material, Part C.

## Results

Q4GOM in noninvasive biological experiments

To demonstrate the optical performance and application potential of the developed technique, Q4GOM was used for quantitative imaging of different types of cells including human cheek cells, blood smear and spontaneously transformed rat embryonic fibroblast (LW3K12) cells. Experimental data were obtained in the setup composed of the add-on 4G optics module connected to the interference microscope (Fig. 1a) using the polarization adapted Mirau MO (Fig. 1b). Details regarding the preparation of specimens and cultivation of cells are described in Supplementary material, Part F. The measured holograms were numerically reconstructed and the resulting data subsequently quantified and visualized using commercial software distributed by TESCAN Orsay Holding, a. s. Before starting biological experiments, the calibration measurements were performed to assess the optical performance of the used setup.

*Spatio-temporal precision and ground-truth accuracy of phase restoration*

To assess the temporal stability and the spatial background noise of Q4GOM in biological experiments, the measurements verifying the spatio-temporal precision and the ground-truth accuracy of the phase restoration were carried out. During the evaluation experiments, Q4GOM was operated without any adaptive hardware control or covering box providing constant atmosphere and thermal stability. Quantitative phase images were reconstructed using numerical procedure which is described in[21] and further discussed in Supplementary material, Part D.

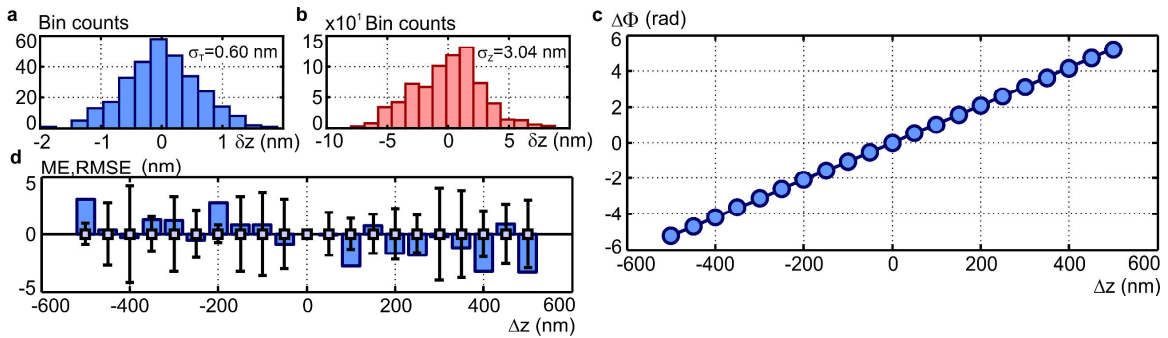

Fig. 2 Evaluation of spatio-temporal precision and ground-truth accuracy of phase restoration in Q4GOM. (a) Histogram demonstrating the temporal stability of the phase imaging (histogram created from 300 reconstructions carried out for a fixed spatial position during 25-minute-long period, reconstructed phase transferred to height variation $\delta z$). (b) Histogram demonstrating the spatial background noise of the phase imaging evaluated in the area of 20x20 µm². (c) Theoretical dependence of the reconstructed phase $\Delta\Phi$ on the ground-truth displacement $\Delta z$ of the piezoelectric transducer (solid line) and the experimentally determined phase $\Delta\Phi$ measured for individual positions $\Delta z$ (circle signs). (d) Accuracy of the restored phase evaluated by the mean error (ME-bars) and the root mean square error (RMSE-error bars) (results were obtained from five independent measurements at each position $\Delta z$).

The temporal stability was evaluated from 300 holograms captured during 25-minute long time period. The reconstructed phase was transferred to heights and the height deviations were evaluated at 9 different parts of the field of view. The size of the evaluated areas corresponded to the diffraction spot of the used MO. A representative height histogram corresponding to the temporal deviations is shown in Fig. 2a. Results obtained in the remaining 8 areas are presented in Supplementary material, Part E. Using the measured height variations, the temporal stability with the standard deviation $\sigma_T = 0.83 \pm 0.23$ nm was obtained. Evaluation of the temporal stability is further documented by Supplementary Movie 1.

The spatial background noise was examined using the height variations in 9 blank areas of 20x20 µm² equally distributed over the field of view. The height histogram representing spatial variations of the restored phase in one selected area is shown in Fig. 2b. Results obtained in the remaining 8 areas are presented in Supplementary material, Part E. Using the spatial height variations, the spatial accuracy given by the standard deviation $\sigma_Z = 2.95 \pm 0.51$ nm was obtained.

In the last calibration measurement, the ground-truth accuracy of the phase restoration was evaluated. In the experiment, the heights retrieved from the phase measurement were compared with the ground-truth heights. The ground true heights were provided by the displacement of the plane mirror loaded on a piezoelectric transducer. In the measurement, the plane mirror was gradually displaced from the initial reference position to the final position 1 µm away with the step of 100 nm. This axial range assures a reliable operation of the piezoelectric transducer and in experiments with biological samples is not exceeded. The same experiment was repeated 5 times to obtain the standard deviation of the independent measurements. The solid line in Fig. 2c shows a theoretical dependence of the restored phase on the mirror displacement, while the circles demonstrate experimental results representing one from 5 measurements. In Fig. 2d, the accuracy is demonstrated by the mean error (ME-bars) and the root mean square error (RMSE-error bars) of 5 independent measurements. The obtained results verify that Q4GOM is able to measure ground-truth heights with the accuracy well below 5 nm.

*Demonstrating versatility of Q4GOM in artifact-free QPI*

The versatility and strengths of Q4GOM are convincingly demonstrated when imaging different types of cells (Fig. 3). Fig. 3a shows the QPI of human cheek cells. The enlarged parts of the field of view in Fig. 3b,c demonstrate the enhanced contrast of the quantitative phase image compared to bright field image. In the QPI, both nucleus and cytoplasmic organelles can be observed and quantitatively characterized. Fig. 3d demonstrates imaging of smaller and more concentrated objects that are represented by human red blood cells. Comparison between the QPI and the bright field image is shown in Fig. 3e,f. The experimental result in Fig. 3g shows the QPI of 100% confluent LW3K12 cells. Comparison between the QPI and the bright field image is presented in Fig. 3h,i. The demonstrations in Fig. 3 illustrate the fact that Q4GOM provides quantitative information in artifact-free high-contrast images without the need of staining, which might possibly affect the behavior of images cells. In Supplementary material, Part G, we further demonstrate the advantages of quantitative nature of the reconstructed images that simplifies numerical processing and automate evaluation of measured data. These options are crucial for high throughput microscopy working with large dataset and implementation of machine learning methods.

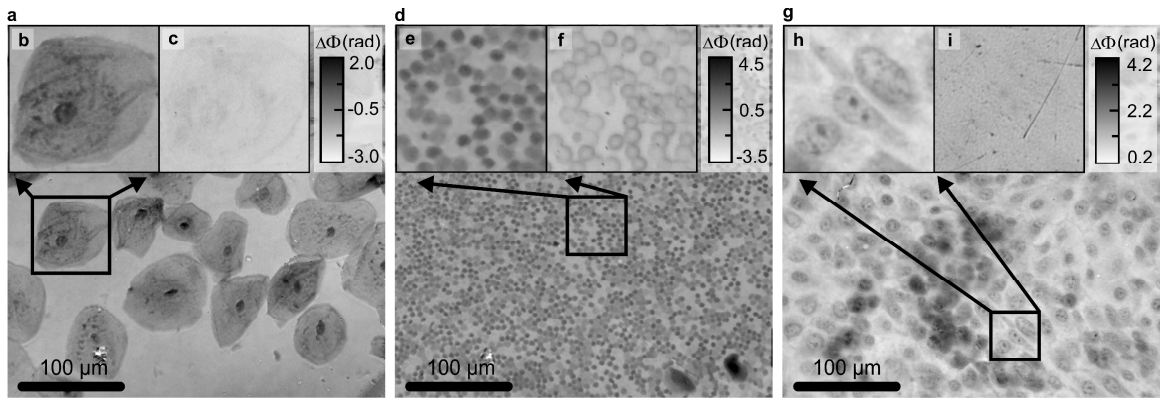

Fig. 3 Demonstration of Q4GOM in the QPI experiments using various cell types. (a) The QPI of human cheek cells. (b),(c) Comparison of the QPI and bright field image of the area marked in (a). (d) The QPI of human blood smear. (e),(f) Comparison of the QPI and bright field image of the area marked in (d). (g) The QPI of 100% confluent LW3K12 cells. (h),(i) Comparison of the QPI and bright field image of the area marked in (g).

*Time-lapse imaging and cell growth study*

To demonstrate the instantaneous operation and the inherent temporal stability of Q4GOM, we performed time-lapse imaging of healthy LW3K12 cells. The cells were imaged for 140-minute-long time period, with the time step 20 seconds. Representative image from the measured time series is presented in Fig. 4a. For subsequent processing, the cells were segmented from the background by watershed segmentation approach and identified as separate regions as shown in Fig. 4b. After the segmentation, area and average dry mass density were extracted from the cells labeled 3 and 5. Both cell parameters were monitored over the whole time period of imaging as represented in Fig. 4c. The area of both cells is increasing over time, while the average density of cell dry mass has a decreasing trend, which is in correspondence with the visual observation. The density of cell dry mass in units of $pg/\mu m^2$ is calculated from the measured phase in radians according to Davies.[22] Representative images of studied cells 3 and 5 along with cell 23 in five time instants with 30 minutes time increment are demonstrated in Fig. 4d. Detail behavior of cells 3 and 5 during the evaluated time extent is shown in Supplementary Movie 2. The experiment demonstrates the potential of Q4GOM for cell tracking and monitoring of cell parameters describing cell behavior in time-lapse images. Only area and dry mass density of the cell were extracted here, however, more extensive list of cell parameters can be obtained.[23]

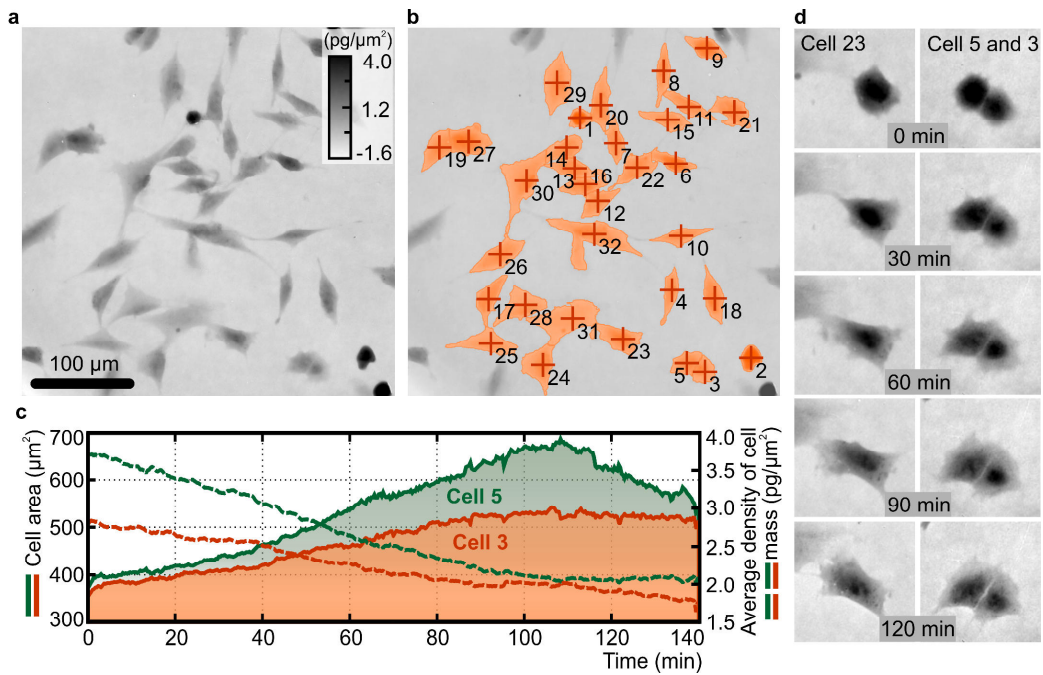

Fig. 4 Time-lapse imaging of live LW3K12 cells and monitoring of area and average density of cell dry mass. (a) Representative image chosen from 140-minute-long experiment. (b) Illustration of cell segmentation used for post-processing of measured data. (c) Quantitative evaluation of cell area and average density of cell dry mass for cell 3 (red color) and 5 (green color). Solid and dashed lines represent areas occupied by cells and average density of cell dry mass, respectively. The measurement was performed with time step 20 seconds. Images are shown in units of pg/µm² recalculated from phase in radians according to Davies.[22] (d) Representative images of selected cells during the measured period.

*Advanced experiments with morphologically changed cells*

The instantaneous ground-truth phase restoration that is in Q4GOM maintained with an exceptional temporal stability and accuracy was successfully deployed also in advanced biological experiments with morphologically changed LW3K12 cells. The gradual morphological change was induced by phosphate-buffered saline (PBS) causing nutritional deprivation, while the more rapid change of morphology was a result of trypsin application. In these experiments, we carried out time-lapse imaging and performed advanced evaluation of measured data including monitoring of cell dry mass density, cell classification based on the morphological parameters and visualization of dynamic dry mass changes. Details regarding the induction of morphological changes of cells are described in Supplementary material, Part F. The results of advanced experiments executed in the automated manner and showing a predestination of Q4GOM for supervised machine learning are available in Supplementary material, Part H, and further documented by Supplementary Movie 3 and Movie 4.

Q4GOM in quantitative retardance imaging of birefringent liquid crystal cells

Multipurpose use of Q4GOM was demonstrated by experiments for quantification of birefringent dynamic-phase retardation that were implemented using the 4G optics module connected to optical microscope with a standard MO (Fig. 1d). In the measurement, liquid crystal display of a reflective SLM (Hamamatsu, X10468-01) was used as a birefringent object. Thanks to the high optical performance of Q4GOM, the phase alterations caused by individual liquid crystal cells were quantified and reliably resolved.

The SLM display was placed in the front focal plane of the MO (10x, NA=0.3) and illuminated by circularly polarized light composed of two orthogonal linear polarization components. As a result of the electrically induced birefringence, one of the polarization components was altered in phase by individual liquid crystal cells to form a signal wave. The orthogonal polarization component was reflected without phase modulation, creating a reference wave. The dynamic phase at the individual liquid crystal cells was determined from the retardance and quantitatively restored using records taken by the 4G optics module. The results obtained are illustrated in Fig. 5. When the SLM was not addressed (i.e., the voltage was not applied to individual pixels), the phase of the sample wave was not spatially modulated, and the background phase shown in Fig. 5a was reconstructed. In the subsequent measurement, a constant phase shift with the stroke π was introduced in the square area of 5x5 pixels. Due to thickness inhomogeneities of the liquid crystal layer and its local heating or non-uniform electric drive scheme, the SLM operation suffers from a spatially varying phase response. To eliminate this error, several calibration methods were developed that allow converting the desired phase values at individual pixels to the correct pixel addressing values.[24] Q4GOM supports these techniques by direct widefield measurement of real phase variations in individual liquid crystal cells. This capability is clearly demonstrated by experimental results shown in Fig. 5b-d. In Fig. 5b, the restored background phase related to the zero-phase level setting is shown along with the phase restored in the rectangular area containing 5x5 pixels. Although all the pixels were addressed for the phase shift π, the reconstructed phase is different at individual pixels. The phase level in the upper row and in the left column is obviously reduced compared to the desired phase stroke π. This is apparent from the gray-level phase map in Fig. 5b and also from the phase profiles taken along the dashed lines in Fig 5b and shown in Fig. 5c. Fig. 5d illustrates the histograms for the neighboring pixels marked I and II in Fig. 5b. The mean value of the pixel II is very close to the desired phase stroke (μ=3.09) while the mean value of the pixel I slightly deviates (μ=2.85). Since the histogram distributions are not significantly overlapped, the phase of the neighboring pixels is safely resolved for the determined mean values and variances. As such Q4GOM was proven as a powerful tool for widefield incoherent QPI of individual SLM pixels allowing assessing their inhomogeneous phase response and crosstalks. This is a progress over the available techniques that gain phase from the intensity of diffracted light or from interference fringes created by beams originating from reference and phase shifted areas of the SLM.[24,25] Direct access to the phase of the individual SLM pixels provided by Q4GOM is not possible in these methods.

In the same configuration, Q4GOM can be applied to quantitative imaging of birefringence in biological samples,[26,27,28] flow of polymer solutions,[29,30] or stress measurements in either biological specimens[31] or optical materials.[32]

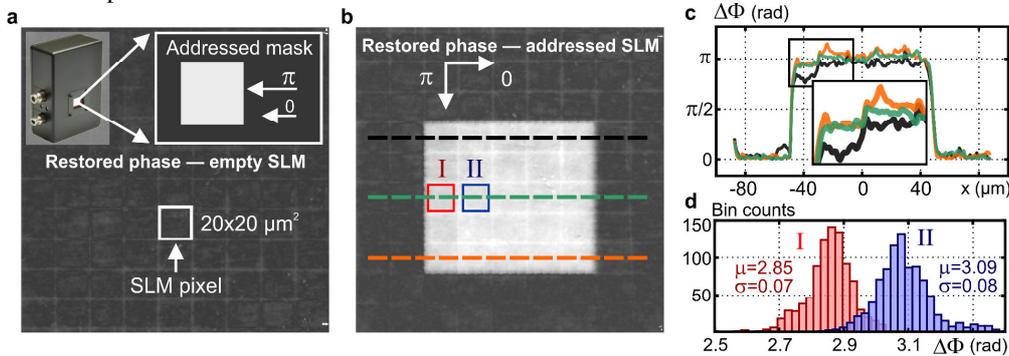

Fig. 5 Q4GOM in quantitative retardance imaging of birefringent liquid crystal cells of a SLM. (a) Illustration of a square phase mask (top) and the phase image of non-addressed SLM. (b) The quantitative phase image of the square phase mask displayed on the SLM (phase stroke π, size 5x5 pixels). (c) Phase profiles taken along the dashed line cross sections in (b). (d) The histograms for pixels I and II in (b).

Q4GOM in quantitative retardance imaging of multi-functional geometric-phase elements

The add-on 4G optics module connected to optical microscope with a standard MO can also be deployed for quantifying the geometric phase of multi-functional optical elements. These elements control light through modulation of the geometric phase and are most often implemented as plasmonic metasurfaces or polarization directed flat elements based on polymer liquid crystals. In prove of principle experiment, we deployed Q4GOM for the quantitative restoration of spatial variations of the geometric phase introduced by a polarization directed flat lens.[4,5] The experiment was carried out using a low magnification MO (4x, NA=0.1) providing the field of view of 0.5x0.5 mm². The measured geometric-phase lens (f=100 mm) was illuminated by the linearly polarized light composed of RHCP and LHCP components as shown in Fig. 1e. The illumination was implemented in Köhler system (Fig. 1a) using a tungsten-halogen lamp with bandpass filter (central wavelength 600 nm, full width at half maximum 50 nm). In the transformation of light by the geometric-phase lens, the handedness of the circular polarization turns from RHCP to LHCP, and vice versa. Simultaneously with the polarization transformation, the modulation of the geometric phase is carried out. The change of the geometric phase is polarization selective and its spatial variations are determined by the local angular rotation of the anisotropy axis of the polymer crystals. In this way, a quadratic phase profile of converging and diverging lens is created for incident waves with RHCP and LHCP, respectively. At individual points of the polarization directed flat lens with the coordinates $x,y$, the phase $\varphi(x,y)$ and $-\varphi(x,y)$ was imposed on the sample and reference waves with the orthogonal circular polarizations (Fig. 1e). Hence, the phase $2\varphi(x,y)$ was restored from the off-axis hologram recorded in the Q4GOM setup.

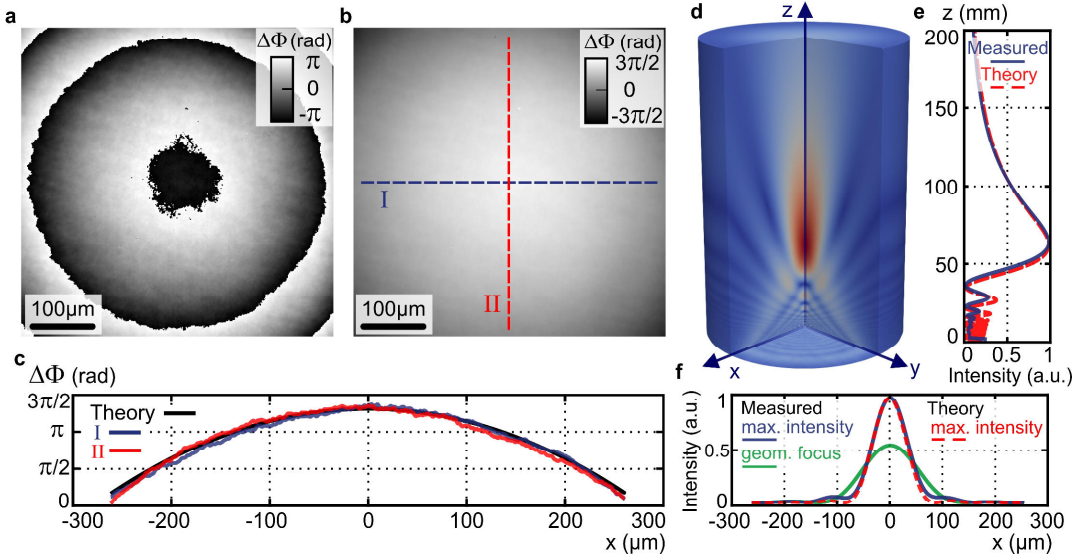

Fig. 6 Q4GOM in quantitative restoration of the geometric phase of a polarization directed flat lens. (a) Restored raw geometric phase mapping phase variations at the plane of the polarization directed flat lens. (b) Unwrapped phase from (a). (c) Phase profiles along the cross sections I (blue) and II (red) compared with theoretical phase profile (black) created for the focal length f=100 mm. (d) 3D amplitude distribution of the focused field calculated from the reconstructed geometric phase. (e) Experimental axial intensity obtained from 3D amplitude distribution in (d) (blue) and theoretical intensity profile for diffraction limited lens (red). (f) Experimental (blue) and theoretical (red) transverse intensity profiles at the plane of maximum axial intensity and experimental transverse intensity profile at the paraxial focal plane (green).

The raw and unwrapped phase images of the measured polarization directed flat lens are shown in Fig. 6a,b. The reconstructed phase $\Delta\Phi(x,y)$ reveals a spherical wavefront $\varphi(x,y)$ shaped by the polarization directed flat lens. To assess the measurement, the phase profiles were taken along dashed line cross sections I and II and compared with the theoretical phase profile created for the focal length of the lens f=100 mm (Fig. 6c). The difference between the theoretical and measured phase profiles was below 0.14 rad in the whole field of view. To demonstrate potential of the measurement for characterization of the geometric-phase elements, we used the measured geometric phase $\Delta\Phi(x,y)$ for numerical calculation of the light propagation behind the lens, which allowed to examine the focused field. The 3D amplitude distribution of the light behind the polarization directed flat lens is visualized in Fig. 6d. This amplitude distribution was obtained as a numerical solution of the Kirchhoff diffraction integral.[33] From the amplitude of the focused field, we calculated the axial intensity response shown in Fig. 6e. The axial intensity is significantly asymmetrical with rapid oscillations in the near distance from the lens and a gradual decrease in the distant region. Furthermore, the intensity maximum is not located at the paraxial focal plane at the distance of 100 mm from the lens but is significantly shifted toward the lens. This specific profile of the axial intensity is a result of the focusing performed with a low Fresnel number,[34] which is given by $N_F = 1$ in the demonstrated case. Using the Fresnel approximation, the relation describing the axial intensity of diffraction limited lens can be obtained.[35] In Fig. 6e, the theoretical axial intensity of an ideal lens (red dashed line) is compared with the intensity profile obtained from measured data (blue line). Both axial intensity profiles are matched well and have a perfect coincidence of intensity maxima and zero points. The transverse intensity profile of the focused field taken at the position of the maximum axial intensity is shown in Fig. 6f (blue line) together with the theoretical profile for the diffraction limited lens (red dashed line). The experimental transverse intensity corresponding to the paraxial focal plane is illustrated by green line in Fig. 6f.

The Q4GOM configuration used in this experiment can also be applied to the measurement of geometric-phase metasurfaces[36] and optically active compounds or biological specimens.

## Discussion

Taking full advantage of the emerging 4G optics technology, we have developed multipurpose quantitative phase retardance microscopy, referenced here as Q4GOM, in which information on the spatially varying ground-truth phase is acquired through the retardance quantification supported by the geometric-phase control of light. The multipurpose use of the method lies in providing spatially resolved restoration of the phase retardance introduced either in the dynamic or geometric phase by isotropic, birefringent and chiral samples or multi-functional optical elements. The widefield quantitative phase images are reconstructed by instantaneous (single-shot) processing of holographic records taken in inherently stable common-path setup using broadband spatially incoherent light. The optical performance and versatility of Q4GOM was demonstrated by its deploying in advanced noninvasive live cell imaging, quantitative imaging of electrically induced birefringent retardance of liquid crystals and mapping of the geometric phase of polarization directed lenses based on a spatially variable rotation of the anisotropy axis in polymer liquid crystals.

As such, Q4GOM brings a new concept for the QPI in traditional areas, such as biophotonics and optical topography, where it may execute imaging of cells with rapid change of morphology or topographic or stress measurement of dynamic events. Progress also lies in subnanometer precision and excellent ground-truth accuracy (well below 5 nm) guaranteed in simple to implement experiments that can be operated under common conditions. The use of Q4GOM goes far beyond the demonstrated experiments and opens new pathways to develop the quantitative retardance imaging of anisotropic biological samples[26,27,28] and flow of polymer solutions,[29,30] to implement stress measurements[31,32] and to provide new tools for the calibration of liquid crystal devices.[24,25] Another significant area in which Q4GOM excels is the quantitative phase imaging of multi-functional optical elements that control light through the geometric phase and are collectively referred to as geometric-phase holograms.[5,6] In addition to the demonstrated measurement of polarization directed 4G optics elements, Q4GOM has also been successfully deployed for the QPI of plasmonic metasurfaces.[36] Thanks to the unique optical performance of Q4GOM, the QPI of the plasmonic metasurfaces with the sensitivity down to a single nanoantenna has been demonstrated in pilot experiments. These experiments open new research directions in the design, optimization and diagnostics of optical composite nanostructures and support their application to biomolecular sensing.

## Funding

Grant agency of the Czech Republic (18-01396S); Ministry of Education, Youth and Sports of the Czech Republic (LQ1601, CEITEC 2020). PB has been supported by scholarship awarded by the Czechoslovak Microscopy Society.

## Author's contributions

P.B., Z.B. and R.Ch. proposed principle of quantitative measurement, P.B. performed experiments and L.Š. processed the experimental data. Z.D. realized polarization modification of Mirau interference objective and assisted with experiments. Z.B. contributed with theoretical model and numerical simulations of the measurement. P.B. and Z.B. wrote the manuscript. R. Ch and Z.B. directed and supervised the project. All authors discussed the results and commented on the manuscript.

## Competing interests

P. B., R. Ch., and Z. B. report a Czech patent, number 307520, and pending PCT application for the Quantitative 4G optics microscope, filed by Brno University of Technology and Palacký University Olomouc.

**References**

1. Cuche, E., Bevilacqua, F. & Depeursinge, C. Digital holography for quantitative phase-contrast imaging. *Opt. Lett.* **24,** 291 (1999).
2. Marquet, P. *et al.* Digital holographic microscopy: a noninvasive contrast imaging technique allowing quantitative visualization of living cells with subwavelength axial accuracy. *Opt. Lett.* **30,** 468 (2005).
3. de Groot, P. Principles of interference microscopy for the measurement of surface topography. *Adv. Opt. Photonics* **7,** 1 (2015).
4. Kim, J. *et al.* Fabrication of ideal geometric-phase holograms with arbitrary wavefronts. *Optica* **2,** 958–964 (2015).
5. Escuti, M. J., Kim, J. & Kudenov, M. W. Geometric-Phase Holograms. *Opt. Photonics News* **27,** 22–29 (2016).
6. Lee, Y.-H. *et al.* Recent progress in Pancharatnam–Berry phase optical elements and the applications for virtual/augmented realities. *Opt. Data Process. Storage* **3,** 79–88 (2017).
7. Park, Y., Depeursinge, C. & Popescu, G. Quantitative phase imaging in biomedicine. *Nat. Photonics* **12,** 578–589 (2018).
8. Marquet, P., Depeursinge, C. & Magistretti, P. J. Review of quantitative phase-digital holographic microscopy: promising novel imaging technique to resolve neuronal network activity and identify cellular biomarkers of psychiatric disorders. *Neurophotonics* **1,** 020901 (2014).
9. Majeed, H. *et al.* Quantitative phase imaging for medical diagnosis. *J. Biophotonics* **10,** 177–205 (2017).
10. Slabý, T. *et al.* Off-axis setup taking full advantage of incoherent illumination in coherence-controlled holographic microscope. *Opt. Express* **21,** 14747 (2013).
11. Cotte, Y. *et al.* Marker-free phase nanoscopy. *Nat. Photonics* **7,** 113–117 (2013).
12. Kim, T. *et al.* White-light diffraction tomography of unlabelled live cells. *Nat. Photonics* **8,** 256–263 (2014).
13. Paganin, D. & Nugent, K. A. Noninterferometric Phase Imaging with Partially Coherent Light. *Phys. Rev. Lett.* **80,** 2586–2589 (1998).
14. Zheng, G., Horstmeyer, R. & Yang, C. Wide-field, high-resolution Fourier ptychographic microscopy. *Nat. Photonics* **7,** 739–745 (2013).
15. Wang, Z. *et al.* Spatial light interference microscopy (SLIM). *Opt. Express* **19,** 1016 (2011).
16. Popescu, G. *et al.* Fourier phase microscopy for investigation of biological structures and dynamics. *Opt. Lett.* **29,** 2503 (2004).
17. Godden, T. M., Muñiz-Piniella, A., Claverley, J. D., Yacoot, A. & Humphry, M. J. Phase calibration target for quantitative phase imaging with ptychography. *Opt. Express* **24,** 7679 (2016).
18. Lee, K. *et al.* Quantitative Phase Imaging Techniques for the Study of Cell Pathophysiology: From Principles to Applications. *Sensors* **13,** 4170–4191 (2013).
19. Pancharatnam, S. Generalized theory of interference, and its applications. *Proc. Indian Acad. Sci. - Sect. A* **44,** 247–262 (1956).
20. Berry, M. V. Quantal Phase Factors Accompanying Adiabatic Changes. *Proc. R. Soc. A Math. Phys. Eng. Sci.* **392,** 45–57 (1984).
21. Zikmund, T. *et al.* Sequential processing of quantitative phase images for the study of cell behaviour in real-time digital holographic microscopy. *J. Microsc.* **256,** 117–125 (2014).
22. Davies, H. G. & Wilkins, M. H. F. Interference Microscopy and Mass Determination. *Nature* **169,** 541–541 (1952).
23. Strbkova, L., Zicha, D., Vesely, P. & Chmelik, R. Automated classification of cell morphology by coherence-controlled holographic microscopy. *J. Biomed. Opt.* **22,** 1 (2017).
24. Engström, D., Persson, M., Bengtsson, J. & Goksör, M. Calibration of spatial light modulators suffering from spatially varying phase response. *Opt. Express* **21,** 16086 (2013).
25. Reichelt, S. Spatially resolved phase-response calibration of liquid-crystal-based spatial light modulators. *Appl. Opt.* **52,** 2610 (2013).
26. Aknoun, S., Bon, P., Savatier, J., Wattellier, B. & Monneret, S. Quantitative retardance imaging of biological samples using quadriwave lateral shearing interferometry. *Opt. Express* **23,** 16383 (2015).

27. de Boer, J. F., Milner, T. E., van Gemert, M. J. C. & Nelson, J. S. Two-dimensional birefringence imaging in biological tissue by polarization-sensitive optical coherence tomography. *Opt. Lett.* **22,** 934 (1997).
28. Shin, I. H., Shin, S.-M. & Kim, D. Y. New, simple theory-based, accurate polarization microscope for birefringence imaging of biological cells. *J. Biomed. Opt.* **15,** 016028 (2010).
29. Haward, S. J., McKinley, G. H. & Shen, A. Q. Elastic instabilities in planar elongational flow of monodisperse polymer solutions. *Sci. Rep.* **6,** 33029 (2016).
30. Haward, S. J., Oliveira, M. S. N., Alves, M. A. & McKinley, G. H. Optimized Cross-Slot Flow Geometry for Microfluidic Extensional Rheometry. *Phys. Rev. Lett.* **109,** 128301 (2012).
31. Sugimura, K., Lenne, P.-F. & Graner, F. Measuring forces and stresses in situ in living tissues. *Development* **143,** 186–196 (2016).
32. McCann, S., Sato, Y., Ogawa, T., Tummala, R. R. & Sitaraman, S. K. Use of Birefringence to Determine Redistribution Layer Stresses to Create Design Guidelines to Prevent Glass Cracking. *IEEE Trans. Device Mater. Reliab.* **17,** 585–592 (2017).
33. Born, M., Wolf, E. & Bhatia, A. B. *Principles of Optics: Electromagnetic Theory of Propagation, Interference and Diffraction of Light*. (Cambridge University Press, 2000).
34. Mahajan, V. N. Axial irradiance and optimum focusing of laser beams. *Appl. Opt.* **22,** 3042 (1983).
35. Běhal, J. & Bouchal, Z. Optimizing three-dimensional point spread function in lensless holographic microscopy. *Opt. Express* **25,** 29026 (2017).
36. Hsiao, H.-H., Chu, C. H. & Tsai, D. P. Fundamentals and Applications of Metasurfaces. *Small Methods* **1,** 1600064 (2017).